\documentclass[12pt]{article}
\usepackage{graphics}

\newcommand{\B}{\beta}
\newcommand{\G}{\gamma}
\newcommand{\Bbb}{B^{}_{\B\B}}
\newcommand{\Bbg}{B^{}_{\B\G}}
\newcommand{\Bgg}{B^{}_{\G\G}}
\newcommand{\Sg}{\sin\!3\G}
\newcommand{\dbe}{\partial_{\B}}
\newcommand{\dga}{\partial_{\G}}
\newcommand{\rnw}{\sqrt{r\over w}}

\begin{document}

\title{ Collective Quadrupole Excitations\\
                in Transitional Nuclei
	  }
\author{
              K. Pomorski, L. Pr\'ochniak and K. Zaj\c ac\\
{\it Institute of Physics, Maria Curie-Sk{\l}odowska University, Lublin, Poland}
\\  
                   S. G. Rohozi\'nski and J. Srebrny\\
{\it Department of Physics, Warsaw University, Warsaw, Poland}}
\maketitle
\date{}
\begin{abstract}
The generalized Bohr Hamiltonian was used to describe the low-lying
collective excitations in even-even isotopes of Ru, Pd, Te, Ba and Nd.
The Strutinsky collective potential and cranking inertial functions 
were obtained using the Nilsson potential. 
The effect of coupling with the pairing vibrations is taken into
account approximately when determining the inertial functions. The
calculation does not contain any free parameter.  
\end{abstract}

\section{Introduction}

For a long time the generalized Bohr hamiltonian (GBH) [1-3] was used to
describe the low lying quadrupole collective excitations in nuclei.
Especially the Bohr hamiltonian with the collective inertial
functions and potential evaluated microscopically (see e.g. [2,3]) was
attractive as a model containing no free parameters. Unfortunately
confrontation of theoretical predictions of such a model with the
experimental data leads to the conclusion that the microscopic inertial
functions, i.e. mass parameters and moments of inertia, are too 
small. One has to magnify them 2 to 3 times in order to obtain the
collective energy levels in right positions [2]. In paper [3] it was
suggested that the pairing correlations in the collective excited states
are weaker than in the ground state. This effect could explain the growth
of the inertial functions. It was shown in Ref. [3] that decrease of
the pairing strength by only 17\%
could increase the magnitude of the mass parameters by a factor 2 to 3 and
in consequence obtain the energies of collective levels relatively
close to the experimental data.

A nice explanation of the origin of the decrease of the pairing correlations
in the collective excited states offers the collective pairing hamiltonian
first introduced by B\`es and coworkers in Ref. [4] for the two--levels model
and than elaborated in [5] for a more realistic case. It was shown in [5]
that the growth of the mass parameter with decreasing pairing
gap ($\Delta$) produces a significant collective effect, namely that the most
probable $\Delta$ is smaller than that obtained from the BCS solution.
The coupling of the collective pairing vibrations with the collective
quadrupole excitations was discussed in Ref. [6] for the axially
symmetric case. The spectrum of collective levels obtained in the model
with the coupling was almost twice compressed in comparison with the spectrum
given by the Bohr hamiltonian which does not contain the coupling with 
pairing vibrations.
Encouraging by the results obtained in  [6] we have modify in Refs. [7,8]
the generalized Bohr hamiltonian taking into account the major effect of the
coupling with the pairing vibrations. Namely, we have evaluated (in each
$\beta,\gamma$ point) all inertial functions for the most probable $\Delta$
not for that which corresponds to the BCS minimum.

In the present paper we are going to describe briefly our model and 
present some typical results for the neutron--rich isotopes of Pd and Ru
and the neutron--deficient isotopes of Te, Ba and Nd.
These examples illustrate well, how does work the model for transitional
nuclei.

\section{The model}

It is rather difficult to solve the nine dimensional eigenproblem of the
full collective hamiltonian containing quadrupole and pairing vibrations
for neutrons and protons. But assuming that the coupling between
quadrupole and pairing variables is weak one can neglect mixing terms
and obtain an approximate solution. Such approximate collective
hamiltonian consists of two known terms and an operator $\hat{\cal H}_{\rm
int}$ which mix quadrupole and pairing variables:
\begin{equation}
\hat{\cal H}_{\rm CQP}  = \hat{\cal H}_{\rm CQ}(\B,\G,\Omega; \Delta
^p,\Delta^n) + \hat{\cal H}_{\rm CP}(\Delta^p,\Delta^n; \B,\G)
+ \hat{\cal H}_{\rm int}\,\,.
\end{equation}
The last term will be neglected in further calculations.
The operator $\hat{\cal H}_{\rm CQ}$ describes quadrupole
oscillations and rotations of a nucleus and it takes the
form of the generalized Bohr hamiltonian [2,3]:
\begin{equation}
\hat{\cal H}_{\rm CQ} = \hat{\cal T}_{\rm vib}(\B ,\G;\Delta^p,\Delta^n)
+ \hat{\cal T}_{\rm rot}(\B ,\G ,\Omega;\Delta^p,\Delta^n) + V_{\rm
coll}(\B,\G;\Delta^p,\Delta^n) \,\,.
\end{equation}
Here $V_{\rm coll}$ is the collective potential, the kinetic
vibrational energy reads
\begin{eqnarray}
\nonumber
\hat{\cal T}_{\rm vib}=-{{\hbar}^2\over{2\sqrt{wr}}}\bigg\{
{1\over \B^4}\bigg[ \dbe\bigg( \B^4\rnw \Bgg\dbe\bigg) -
\dbe \bigg(\B^3\rnw\Bbg\dga\bigg)\bigg]+ &&\\
+ {1\over \B\Sg}\bigg[ -\dga \bigg( \rnw\Sg\Bbg\dbe\bigg) +
{1\over\B}\dga \bigg(\rnw\Sg\Bbb\bigg)\dga\bigg]\bigg\}
\end{eqnarray}
and the rotational energy is 
\begin{equation}
\hat{\cal T}_{\rm rot}={1\over 2}\sum_{k=1}^{3} \hat{I}^2_k/{\cal J}_k \,\,.
\end{equation}
The intrinsic components of the total angular momentum are denoted as
$\hat{I}_k,\, (k=1,2,3)$, while $w$ and $r$ are the determinants of the
vibrational and rotational mass tensors. The mass parameters (or
vibrational inertial functions) $\Bbb$, $\Bbg$ and $\Bgg$ together with
moments of inertia ${\cal J}_k,\, (k=1,2,3)$ depend on
intrinsic variables $\B,\G$ and pairing gap values $\Delta^p,\Delta^n$.
All inertial functions are determined from a microscopic theory. We
apply the standard cranking method to evaluate the inertial functions
assuming that the nucleus is a system of nucleons moving in the deformed
mean field (Nilsson potential) and interacting through monopole pairing
forces. 
One has to stress that for $\Delta$ corresponding to the minimum of the
BCS energy the operator $\hat{\cal H}_{\rm CQ}$ is exactly the same as the Bohr
hamiltonian used in Ref. [2,3].

For a given nucleus the second term in Eq. (1) describes collective
pairing vibrations of systems of $Z$ protons and $A-Z$ neutrons
\begin{equation}
\hat{\cal H}_{\rm CP} = \hat{\cal H}^Z_{\rm CP} +
\hat{\cal H}^{A-Z}_{\rm CP}
\end{equation}
and it can be expressed in the following form [4,5]:
\begin{equation}
 \hat{\cal H}^{\cal N}_{\rm CP}=-\frac{\hbar^2}
{2\sqrt{g(\Delta)}}\frac{\partial}
{\partial\Delta}\frac{\sqrt{g(\Delta)}}
{B_{\Delta\Delta}(\Delta)}
\frac{\partial}{\partial\Delta} + V_{\rm pair}(\Delta),
\end{equation}
where ${\cal N}=Z$, $\Delta=\Delta^p$ for protons and, respectively,
${\cal N}=A-Z$, $\Delta=\Delta^n$ for neutrons. The functions appearing
in the hamiltonian (6), namely the pairing mass parameter
$B_{\Delta\Delta}(\Delta)$, the determinant of the metric tensor
$g(\Delta)$ and the collective pairing potential $V_{\rm pair}(\Delta)$
are determined microscopically.
 
Solving the eigenproblem of the collective pairing hamiltonian
(6) one can find the pairing vibrational ground-state wave
function $\Psi_0$ and the ground-energy $E_0$ at each
deformation point. The most probable value of the energy gap $\Delta_{vib}$
corresponds to the maximum of the probability  of finding a given gap
value in the collective pairing ground-state (namely the maximum of the
function $g(\Delta )|\Psi_{0}(\Delta )|^2$). As it is shown in
Fig.~1 the $\Delta_{vib}$ is shifted towards smaller gaps from the
equilibrium point $\Delta _{eq}$ determined by the minimum of $V_{\rm
pair}$ (or by the BCS formalism). Such a behavior of the pairing ground
state function $\Psi_0$ is due to the rapid increase of
pairing mass parameter $B_{\Delta \Delta}$.
In general the ratio of $\Delta _{vib}$ to $\Delta _{eq}$ is of about
$0.7$.

All collective functions appearing in Eqs. (3,4) are
calculated using the most probable pairing gap values for protons and
for neutrons instead the equilibrium ones. The collective potential corresponds
to the ground state of the $\hat{\cal H}_{CP}$ hamiltonian (5) and it is very
close to the BCS energy in each $\beta,\gamma$ point.

The approximation described above is rather crude but it includes 
the main effect (at least on
average) of the coupling with the pairing vibrational mode. This
procedure improves significantly the accuracy in reproducing the
experimental data and it introduces no additional parameters into
the model. Our calculations were done using the standard Nilsson single
particle potential with the shell dependent parametrization found in
Ref. [9]. The pairing strength was fitted in Refs. [7,8] to the mass
differences.
 
\section{Results}

We present here only some examples of results for the neutron--rich isotopes
of Ru and Pd and for three chains of isotopes (Te, Ba and Nd) from the
neutron--deficient region of nuclei. 

In order to illustrate the effect
of the coupling of the quadrupole and pairing vibrations we have
compared in Fig. 2 the energy levels obtained with the traditional
GBH ("old") with those evaluated within the present model ("new").
 As one can learn from Fig.~2
the improvement in reproducing the experimental data caused by
coupling with the pairing vibrations is really significant.

In Fig.~3 we present the theoretical (open symbols) and experimental 
(full symbols) energy levels of the ground state band and the $\gamma$ band
for the  even--even isotopes of Ru and Pd with $64 \leq N \leq 74$ neutrons. 
The agreement of theoretical predictions with the experimental data is
here rather good. The situation in the neutron--deficient nuclei is not
so optimistic. A typical sets of results is presented in Fig.~4, where
the lowest levels of the ground state band of Te, Ba and Nd isotopes are
plotted as a function of the neutron number.  It is seen that a good
agreement is obtained for the Nd nuclei and for the lightest Te and Ba isotopes
only.

The electromagnetic transitions between the band members and between the
states belonging to different bands are also relatively well reproduced
by our model (see Refs. [7,8]).
   
All experimental data in Figs. 2-4 are taken from Ref. [10].
       
\section{Summary and conclusions}

The generalized Bohr Hamiltonian (GBH) [1-3] is used to describe the  low-lying
collective excitations in even-even isotopes of the neutron--deficient and 
neutron--rich regions of nuclei [7,8].
The collective potential and inertial functions are determined by means of
the Strutinsky method and the cranking model, respectively. A shell-dependent
parametrization of the Nilsson potential is used.  There are no
adjustable parameters in the calculation. The coupling of the
quadrupole and pairing vibrations [5-6] is taken into account and it brings 
the energy levels down to the scale comparable with that characteristic for 
the experimental levels [6]. 

In the neutron--reach region we have performed calculations for chains of 
isotopes of Ru and Pd. In this case theoretical estimates of energies of low 
lying collective states and electromagnetic transitions within bands, as well
as between members of different bands, are even closer to experimental data
than for nuclei from the neutron-deficient region. 
In the region of neutron--deficient nuclei
the GBH works better in the case of elements with larger $Z$, namely,
Ce, Nd and Sm than for Xe, Ba and Te which have only two
protons outside the closed shell $Z=50$. Energies are especially well
reproduced by the calculation for isotopes with lower number of neutrons. For
those with neutron number $N=78,\, 80$ the energy levels are, as a rule, too
high. On the contrary, electromagnetic properties seem to be better reproduced
just in the case of heavier isotopes. 

Concluding, we may say that adding of the coupling with the pairing vibrations
to the generalized Bohr Hamiltonian improves significantly the quality of
theoretical estimates for nuclei from the both regions.  

\vspace{5mm} 
\parindent=0cm

\section*{Acknowledgement}
This work was supported in part by the Polish Committee for Scientific 
Research under Contract No. 2~P03B~068~13.

\vspace{2cm}

{\bf Figure captions:}\\
\vspace{0.5cm}

{\bf Fig.~1} The pairing vibration mass parameter ($B_{\Delta\Delta}$),
and potential ($V_{\rm pair}$), and the ground-state function ($\Psi ^N_0$)
as function of the pairing energy gap $\Delta$ for the system of
$60$ neutrons at the deformation point $\beta = 0.2,\, \gamma = 20^{\circ}$.
The equilibrium value of the energy gap is $\Delta_{eq}
\approx 0.14\hbar\omega_0$, the most
probable one is $\Delta_{vib}\approx 0.09\hbar\omega_0$. \\

{\bf Fig.~2} The lowest experimental and the theoretical
(connected by straight lines) excited levels in $^{104}$Ru versus angular
momentum $J^{\pi}$. The theoretical values were calculated including
the effect of coupling with the pairing vibrations ("new") and
without this coupling, i.e. within usual microscopic Bohr model ("old").\\

{\bf Fig.~3} The lowest theoretical and experimental energy levels of the
ground state band and the $\gamma$ band for the chains of Ru and Pd isotopes.\\

{\bf Fig.~4} The lowest theoretical and experimental energy levels of the
ground state band for the chains of Te, Ba and Nd isotopes.

\newpage


\vspace{2cm}
{\includegraphics*{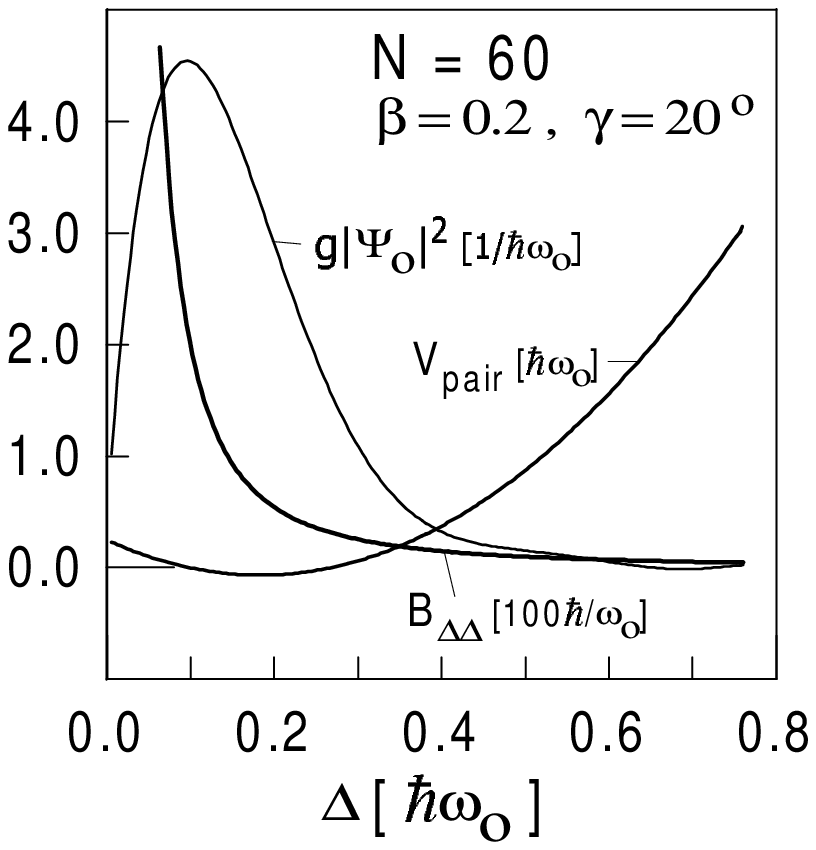}}

\newpage


\vspace{2cm}
{\includegraphics*{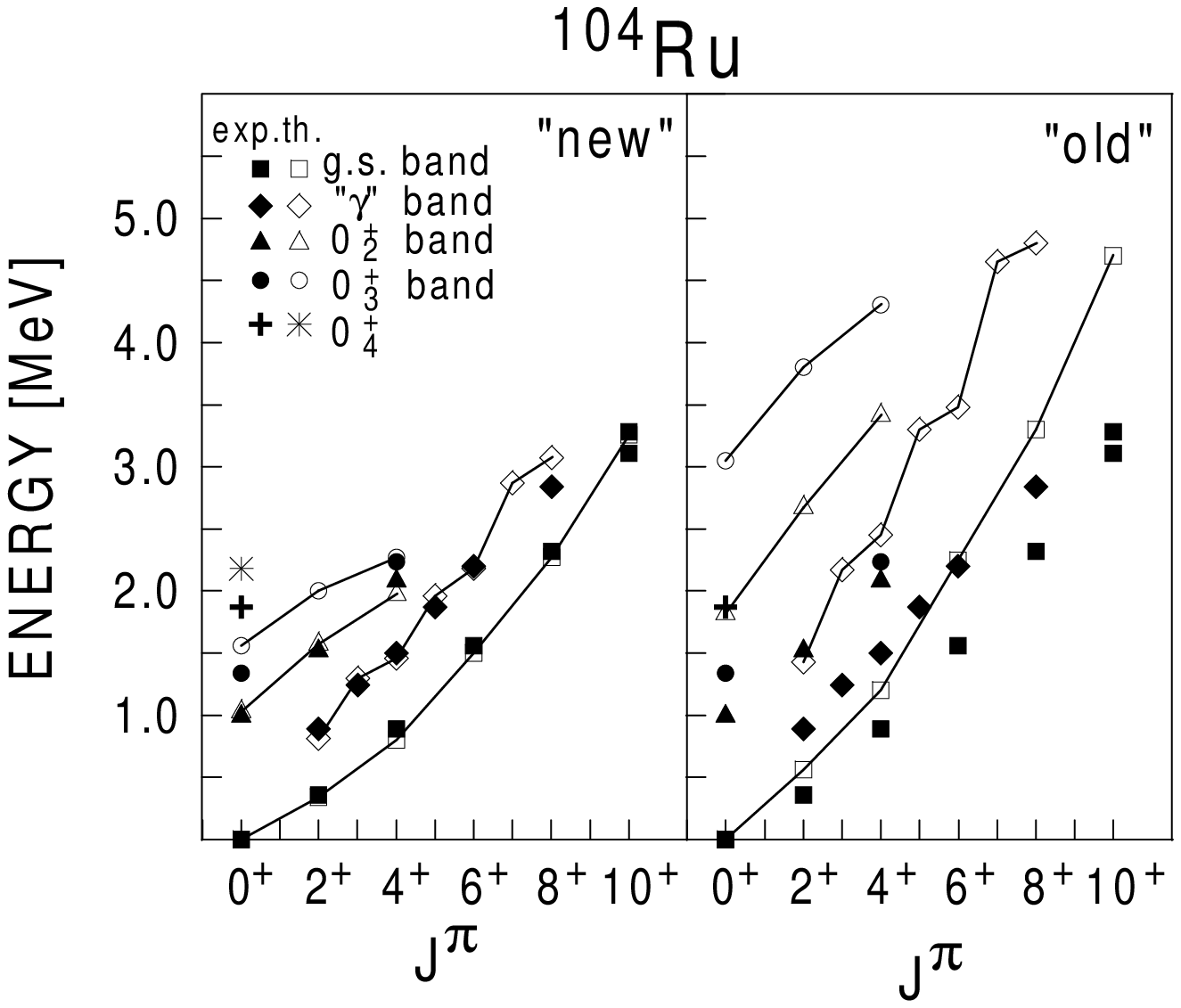}}
\newpage


{\includegraphics*{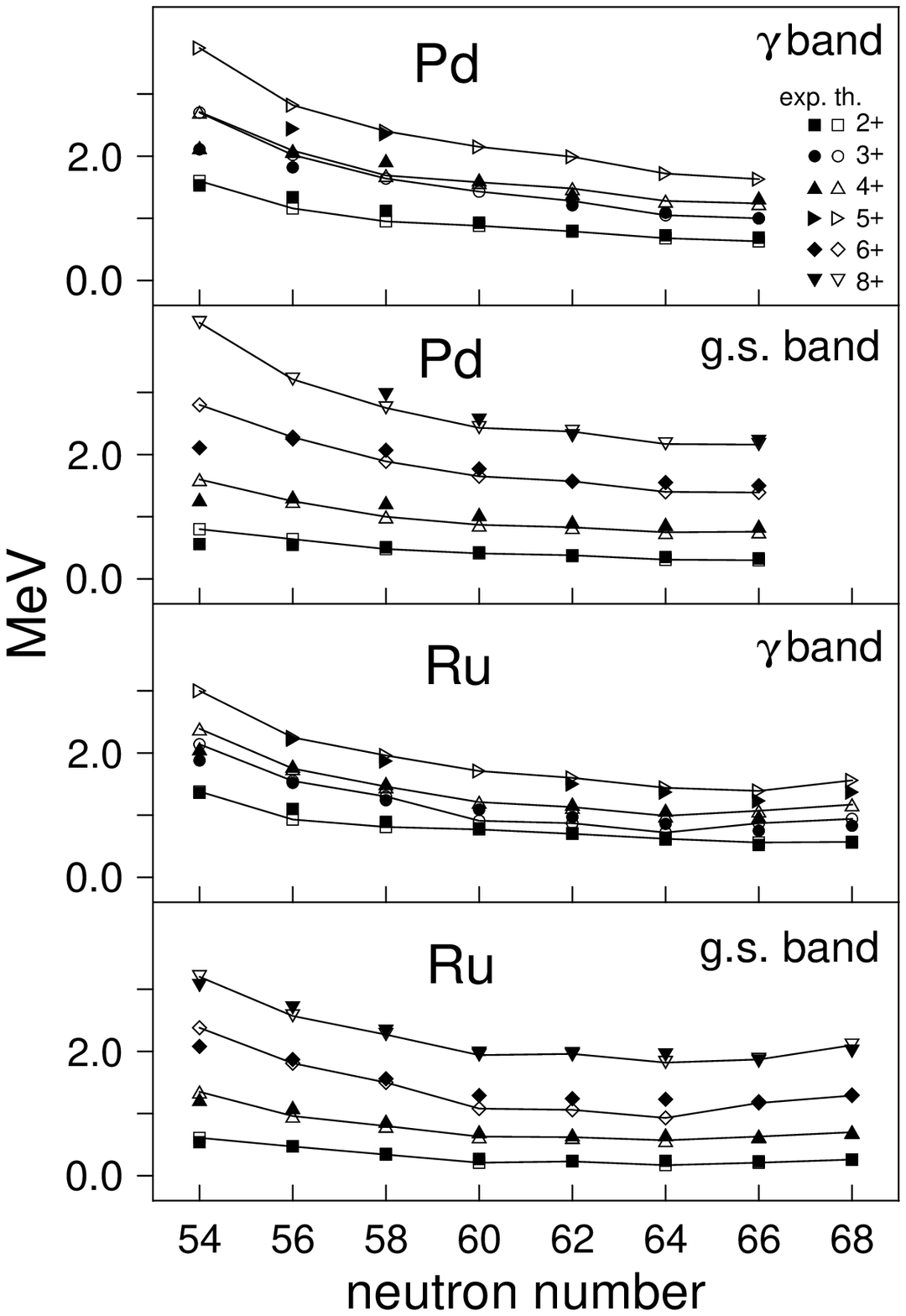}}
\newpage

{\includegraphics*{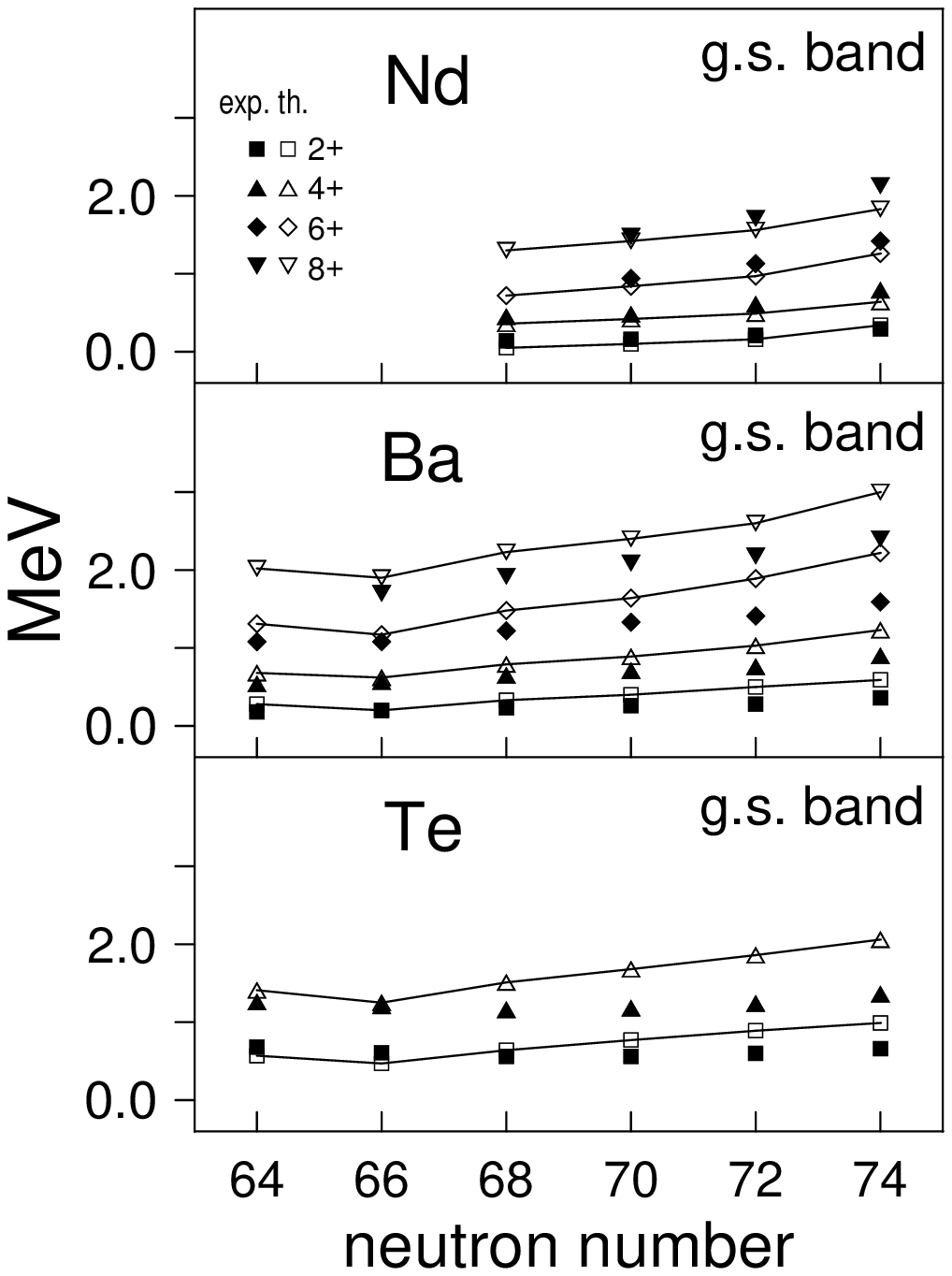}}

\end{document}